\documentclass[aps,pra, article, amsmath, amssymb, showpacs, superscriptaddress]{revtex4}

\usepackage{graphicx}% Include figure files
\usepackage{bm}% bold math
\begin{document}

\title{Cosmological Gravimetry Using High-Precision Atomic Clocks}

\author{V. I. Yudin}
\email{viyudin@mail.ru}
\affiliation{Novosibirsk State University, ul. Pirogova 2, Novosibirsk, 630090, Russia}
\affiliation{Institute of Laser Physics SB RAS, pr. Akademika Lavrent'eva 13/3, Novosibirsk, 630090, Russia}
\affiliation{Novosibirsk State Technical University, pr. Karla Marksa 20, Novosibirsk, 630073, Russia}
\author{A. V. Taichenachev}
\affiliation{Novosibirsk State University, ul. Pirogova 2,
Novosibirsk, 630090, Russia} \affiliation{Institute of Laser
Physics SB RAS, pr. Akademika Lavrent'eva 13/3, Novosibirsk,
630090, Russia}

%\date{\today}

\begin{abstract}
In this paper, a hypothesis that the cosmological gravitational potential can be measured with the use of high-precision atomic clocks is proposed and substantiated. The consideration is made with the use of a quasi-classical description of the gravitational shift that lies in the frame of nonmetric theories of gravity. It is assumed that the cosmological potential is formed by all matter of the Universe (including dark matter and dark energy) and that it is spatially uniform on planet scales. It is obvious that the cosmological potential, $\Phi_\text{CP}$, is several orders of magnitude greater than Earth's gravitational potential $\varphi_\text{E}$
(where $|\varphi_\text{E}/c^2|\sim 10^{-9}$ on Earth's surface). In our method, the tick rates of identical atomic clocks are compared at two points with different gravitational potentials, i.e. at different heights. In this case, the information on $\Phi_\text{CP}$ is contained in the cosmological correction $\alpha\neq 0$ in the relationship $\Delta\omega/\omega=(1+\alpha)\Delta \varphi/c^2$ between the relative change of the frequencies $\Delta \omega/\omega$ (in atomic clocks) and the difference of the gravitational potential $\Delta \varphi$ at the measurement points. We have estimated the low limit of cosmological correction, $\alpha >10^{-6}$. It is shown that using a modern atomic clock of the optical range it is possible to measure the value of $\alpha$ in earth-based experiments if $|\alpha|>10^{-5}$. The obtained results, in the case of their experimental confirmation, will open up new unique opportunities for the study of the Universe and the testing of various cosmological models. These results will also increase the measurement accuracy in relativistic geodesy, chronometric gravimetry, global navigation systems, and global networks of atomic clocks.
\end{abstract}

\pacs{06.30.Ft, 32.10.-f}

\maketitle
\section*{I. Introduction}

Atomic clocks are currently the most precise physical devices. By now, unprecedented relative uncertainty at a level of $10^{-18}$ has been achieved
\cite{Schioppo_2017}. The problem of attaining a level of $10^{-19}$ is on the agenda. Frequency measurements of such accuracy can have a great impact on the further development of fundamental and applied physics (see, for example, review \cite{Ludlow_2015}). In particular, atomic clocks play a special role in astrophysics and cosmology. In this context, the drift of fundamental constants \cite{Angstmann_2004,Karshenboim_2005,Reinhold_2006,Flambaum_2007,Flambaum_2009,Tilburg_2015} and the detection of dark matter \cite{Derevianko_2014,Stadnik_2016,Roberts_2017}  are of the greatest interest.

In this paper, we develop an idea of cosmological gravitation
measurements with the use of high-precision atomic clocks.
According to the general theory of relativity, the clock tick rate in the gravitational field is slowed. As a result, the
frequency of atomic transition experiences a gravitational redshift depending on the value of the gravitational potential. In
the case of a spatially non-uniform gravitational potential, this effect
leads to the following well-known relationship:
\begin{equation}\label{usual}
\frac{\omega ({\bf r}_1)-\omega ({\bf r}_2)}{\omega ({\bf
r}_1)}=\frac{\varphi({\bf r}_1)-\varphi({\bf r}_2)}{c^2}\,,
\end{equation}
which describes the measured relative difference of the frequencies
$\omega ({\bf r}_1)$ and $\omega ({\bf r}_2)$ for the same clock
transition at two different spatial points ${\bf r}_1$ (observer) and ${\bf
r}_2$ (emitter) with different gravitational potentials, $\varphi({\bf r}_1)$
and $\varphi({\bf r}_2)$ ($c$ is the light speed)
\cite{Chou_Sc_2010,Katori_2016}. In particular, formula
(\ref{usual}) allows using high-precision atomic clocks to
measure the difference of the gravitational potentials for various
Earth's surface points, which can form a basis for
the so-called chronometric geodesy (relativistic geodesy, chronometric leveling) \cite{Vermeer_1983}.

In this paper, we propose and substantiate a hypothesis according to
which the following relationship [which is more exact than Eq.~(\ref{usual})] can be used:
\begin{equation}\label{Dw_gen}
\frac{\omega ({\bf r}_1)-\omega ({\bf r}_2)}{\omega ({\bf r}_1)}=
(1+\alpha)\frac{\varphi({\bf r}_1)-\varphi({\bf r}_2)}{c^2}\,,
\end{equation}
where the parameter $\alpha$ does not depend on the type of atomic
clock and has the meaning of a cosmological correction, which
contains information on the cosmological gravitational action of
the Universe on our planetary system. In particular, this
correction can be interpreted as the value that is proportional
to the cosmological gravitational potential $\Phi_\text{CP}$ at Earth's location point.
We assume that the cosmological potential is formed by all matter of the Universe (including
dark matter and dark energy) and that it is spatially uniform on planet scales.
We also assume that the value of the gravitational-cosmological ``background'' $\Phi_\text{CP}$ is several orders of magnitude greater than the gravitational potential of our planet
$\varphi_\text{E}$ (where $|\varphi_\text{E}/c^2|\sim 10^{-9}$ on Earth's surface).
Thus, Eq.~(\ref{Dw_gen}) can be used as a basis for cosmological gravimetry.

Below we develop quasi-classical approach, which leads to a nonzero cosmological correction $\alpha$ in Eq.~(\ref{Dw_gen}). This approach lies in the frame of nonmetric theories of gravity (e.g., see \cite{Nordtvedt_1975}).

\section*{II. Quasi-classical consideration of chronometric gravimetry}

In this section, we will use the results of methodologically simplest approach (e.g., see in Refs.\cite{Yudin_2017,Wolf_2016}), which explains the gravitational
redshift of arbitrary atomic transition as resulting from the ``mass defect'' (for quantum atomic states) in the presence of the classical (Newtonian)
gravitational potential $\varphi({\bf r})$, that is not within the framework of general relativity.
In particular, the frequency at the point ${\bf r}$ is expressed by the following formula (e.g., see in Ref.\cite{Yudin_2017}):
\begin{equation}\label{omega_r}
\omega({\bf r})=\omega^{}_0\left(1+\frac{\varphi({\bf
r})}{c^2}\right),
\end{equation}
where $\omega_0$ is the unperturbed frequency of the atomic transition (clock transition) in the absence of gravitation. Therefore, we will consider chronometric gravimetry using this ``quasi-classical'' approach, which explains our idea of cosmological gravimetry in the frame of nonmetric theories of gravity. For simplicity, we will analyze the stationary case in the absence of motion of celestial bodies.

First of all, let us show an important feature of the formula (\ref{omega_r}) which leads to the absoluteness of the gravitational potential as a uniquely specified function
$\varphi({\bf r})$ in the entire space.
To prove this, we consider the ratio of two frequencies at two different points, ${\bf r}_1$ and ${\bf r}_2$:
\begin{equation}\label{fix_omega}
\frac{\omega({\bf r}_2)}{\omega({\bf r}_1)}=
1-\frac{1}{1+\varphi({\bf r}_1)/c^2}\frac{\varphi({\bf r}_1)-\varphi({\bf r}_2)}{c^2}\equiv fixed\,,
\end{equation}
which is some fixed value measured experimentally. This seemingly obvious and trivial fact leads, nevertheless, to far-reaching consequences. Indeed, if we consider gravitation only from the viewpoint of gravitational force ${\bf F}_\text{grav}$$=$$-M{\bf \nabla} \varphi({\bf r})$ (where $M$ is the mass of a body), the gravitational potential is determined up to an arbitrary constant $C$, because the transformation $\varphi({\bf r})\rightarrow \varphi({\bf r})+C$
does not affect the force. However, the transformation $\varphi({\bf r})\rightarrow \varphi({\bf r})+C$ is absolutely inadmissible from the viewpoint of  formula (\ref{fix_omega}), because the result becomes dependent on $C$ and an arbitrariness of the constant $C$ causes total uncertainty of the ratio ${\omega({\bf r}_1)}/{\omega({\bf r}_2)}$ in a non-uniform gravitational potential for any atomic transition. This does not agree with the experiments and defies common sense. Thus, formula (\ref{fix_omega}), surprisingly, leads to the fact that the gravitational potential is described by some uniquely defined function $\varphi({\bf r})$ for the Universe. In summary, we believe that the expression (\ref{fix_omega}) can be considered as an independent physical principle, which we call the nonlocal chronometric principle. This phenomenological postulate imposes strong restrictions on the gravitational potential. It is, in essence, a change of the boundary conditions if the function $\varphi({\bf r})$ is considered as the solution of the differential equation for infinite space (i.e., for the Universe).

Now let us consider the expression:
\begin{equation}\label{Dw_class}
\frac{\omega ({\bf r}_1)-\omega ({\bf r}_2)}{\omega ({\bf r}_1)}=1-\frac{\omega({\bf r}_2)}{\omega({\bf r}_1)}=\frac{1}{1+\varphi({\bf r}_1)/c^2}\frac{\varphi({\bf r}_1)-\varphi({\bf r}_2)}{c^2}\,,
\end{equation}
which follows from Eq.~(\ref{omega_r}). Within the framework of Newtonian theory of gravitation, the potential $\varphi({\bf r})$ can be structured as follows:
\begin{eqnarray}\label{U_structure}
&& \varphi({\bf r})=\varphi_\text{loc}({\bf r})+\Phi_\text{CP}({\bf r})\,, \\
&&\varphi_\text{loc}({\bf r})=\varphi_\text{E}({\bf r})+\varphi_\text{S}({\bf r})+\varphi_\text{M}({\bf r})+...\,,\nonumber
\end{eqnarray}
where the local potential $\varphi_\text{loc}({\bf r})$ contains the Newtonian potentials of all bodies of the solar system: the Earth $\varphi_\text{E}({\bf r})$, the Sun $\varphi_\text{S}({\bf r})$, the Moon $\varphi_\text{M}({\bf r})$, etc. The second contribution in Eq.~({\ref{U_structure}}) corresponds to the cosmological potential, $\Phi_\text{CP}({\bf r})$, including the potentials of all other bodies of the Universe (located very far from the Earth), and the gravitational effect of dark matter and dark energy. If we describe experiments on Earth's surface or in the near-Earth space,  the distance between the atomic clock positions, $|{\bf r}_1-{\bf r}_2|$, can be at the level of $1-10^{8}$~m. It is obvious that at such small distances we can absolutely neglect the spatial variation of the cosmological potential $\Phi_\text{CP}({\bf r})$, i.e., $\Phi_\text{CP}({\bf r})= const$ and it can be considered as gravitational-cosmological background. Thus, the potential difference in Eq.~(\ref{Dw_class}) is fully described by the local potential:
\begin{equation}\label{DU_loc}
\varphi({\bf r}_1)-\varphi({\bf r}_2)=\varphi_\text{loc}({\bf r}_1)-\varphi_\text{loc}({\bf r}_2) \approx \varphi_\text{E}({\bf r}_1)-\varphi_\text{E}({\bf r}_2)\,.\nonumber
\end{equation}
Let us estimate the various contributions in Eq.(\ref{U_structure}). Near Earth's surface we have an estimate of the Earth potential $|\varphi_\text{E}/c^2|\approx 0.7\times 10^{-9}$, and
the estimate $|\varphi_\text{S}/c^2|\approx 10^{-8}$ corresponds to the Sun potential in Earth's orbit. For a lower estimate of the cosmological potential $\Phi_\text{CP}$, we consider the orbital motion of the solar system around the center of the Galaxy at a speed $v_S\approx 240$~km/s. Consequently, we have a rough lower estimate for the cosmological potential, $|\Phi_\text{CP}/c^2|>|\varphi_{\rm Galaxy}/c^2|\approx v_S^2/c^2 \sim10^{-6}$. Moreover, because our Galaxy is only an insignificant part of the Universe, we can expect, in reality, a much stronger value, $|\Phi_\text{CP}/c^2|\gg 10^{-6}$. In any case, near Earth's surface the following condition is valid: $|\Phi_\text{CP}|\gg |\varphi_\text{loc}({\bf r})|$. Therefore, for the denominator in the right-hand side of  formula (\ref{Dw_class}) we can use approximation: $\varphi({\bf r}_1)\approx \Phi_\text{CP}$. Thus, taking into account Eq.~(\ref{DU_loc}), the expression (\ref{Dw_class}) can be rewritten in the following form:
\begin{equation}\label{Dw_class_appr}
\frac{\omega ({\bf r}_1)-\omega ({\bf r}_2)}{\omega ({\bf r}_1)}\approx\frac{1}{1+\Phi_\text{CP}/c^2}\frac{\varphi_\text{loc}({\bf r}_1)-\varphi_\text{loc}({\bf r}_2)}{c^2}\,,
\end{equation}
which clearly shows that the cosmological potential $\Phi_\text{CP}$ can be measured using high-precision atomic clocks. Comparing formulas (\ref{Dw_class_appr}) and (\ref{Dw_gen}), we find the following interrelation:
\begin{equation}\label{alpha}
\alpha=-\frac{\Phi_\text{CP}/c^2}{1+\Phi_\text{CP}/c^2}\quad\Leftrightarrow \quad \Phi_\text{CP}/c^2=-\frac{\alpha}{1+\alpha}\,.
\end{equation}
In the case of $|\Phi_\text{CP}/c^2|\ll 1$,  we have the following approximation:
\begin{equation}\label{alpha_approx}
\alpha\approx -\Phi_\text{CP}/c^2\,.
\end{equation}
Because the Newtonian gravitational potential is negative (the attractive potential), the cosmological potential is also negative, $\Phi_\text{CP}({\bf r})<0$ (see comment \cite{comm_C0}). This means that the cosmological correction $\alpha$ in Eq.~(\ref{Dw_gen}) has to be positive (at least within the quasi-classical consideration), $\alpha >0$.

In spite of the fact that $|\Phi_\text{CP}|$$\gg$$|\varphi_\text{loc}|$, the existence of a great constant gravitational-cosmological background $\Phi_\text{CP}$ practically does not affect the relative motion of bodies in the solar system, because mechanical motion is determined by the gravitational force ${\bf F}_\text{grav}$$=$$-M{\bf \nabla} \varphi({\bf r})$, which is invariant relative to the transformation $\varphi({\bf r})\rightarrow \varphi({\bf r})+C$. Thus, in our case the gravitational force describing the relative motion of bodies in the solar system can be considered only as a consequence of the local potential: ${\bf F}_\text{grav}$$=$$-M{\bf \nabla} \varphi_\text{loc}({\bf r})$.

The above reasonings show that chronometric investigations involve more fundamental aspects of gravitation in comparison to study of only mechanical motion of bodies in the gravitational field.
In particular, it is shown that  the concept of cosmological gravitational potential $\Phi_\text{CP}({\bf r})$ taken into account in Eqs.~(\ref{Dw_class}) and (\ref{Dw_class_appr}) is justified, despite a great distance of objects of deep space to the local planetary system. In our opinion, there are no reasons to believe that $\Phi_\text{CP}({\bf r}_\text{E})=0$ (where ${\bf r}_\text{E}$ is the coordinate of Earth's center). Moreover, we have shown [see Eqs.~(\ref{Dw_gen}),(\ref{Dw_class_appr})-(\ref{alpha_approx})] how the value of $\Phi_\text{CP}({\bf r})$ can be experimentally measured in the case of $|\Phi_\text{CP}({\bf r})|\gg |\varphi_\text{loc}({\bf r})|$. In the essence, cosmological correction $\alpha$ in Eq.~(\ref{Dw_gen}) corresponds to the nonlinear theory on the total gravitational potential $\varphi({\bf r})$ [see Eq.~(\ref{Dw_class})].

Summarizing this section, our quasi-classical consideration of chronometric gravimetry (i.e., in the frame of nonmetric theory of gravitation) leads to the following prediction: cosmological correction $\alpha$ is a positive number ($\alpha > 0$) and its value has the low limit
\begin{equation}\label{limit}
\alpha > 10^{-6}\,,
\end{equation}
which arises due to Galaxy's gravitational potential, $|\varphi_{\rm Galaxy}/c^2|\sim 10^{-6}$. Thus, if experiments will show
\begin{equation}\label{cosm}
\alpha > 2\times 10^{-6}\,,
\end{equation}
then it will indicate that the cosmological contribution exceeds the galactic contribution.

\section*{III. Prospects for experimental measurements of the cosmological correction {\large $\alpha$}}
It should be noted that formula of the type (\ref{Dw_gen}) has been used for many decades in the relativistic theory of gravitation and in the processing of some gravimetric experiments (see, for example, review \cite{Will_2014}). However, a hypothetic difference from zero ($\alpha\neq 0$) was interpreted as violation of the principle of local position invariance when $\alpha$ depends on the type (nature) of atomic clock (see Ref.~\cite{Will_2014}). In contrast, we predict universality of formula (\ref{Dw_gen}), which should not be associated only with the violation of local position invariance. In our approach, the parameter $\alpha$ does not depend on the type of atomic clock and contains information on cosmological gravitation.

In this context, it is of interest to analyze some experiments that have been performed for the last several decades (starting from 1960). These results are presented in Fig.~3 in Ref.~\cite{Will_2014}. Note that in this figure we should not take into account so-called ``null redshift experiments'' (denoted by red arrows in Fig.~3 in Ref.~\cite{Will_2014}). In ``null redshift experiments'' it is assumed that $\alpha\neq 0$ [in formula (\ref{Dw_gen})] depends on the type of atomic clocks. Then, comparing the behavior of two different clocks based on different atomic transitions (with frequencies $\omega^{(1)}$ and $\omega^{(2)}$), one can measure with high accuracy the difference between the corresponding coefficients, $(\alpha_1-\alpha_2)$:
\begin{equation}\label{null}
\frac{\Delta\omega^{(1)}}{\omega}-\frac{\Delta\omega^{(2)}}{\omega}=(\alpha_1-\alpha_2)\frac{\Delta \varphi}{c^2}\,.
\end{equation}
The logic of ``null redshift experiments'' is following: if the experiments show that $(\alpha_1-\alpha_2)\rightarrow 0$, this means that $\alpha_{1,2}\rightarrow 0$, that is, the principle of local position invariance is not violated. However, as shown above, the presence of $\alpha\neq 0$ in formula (\ref{Dw_gen}) can be associated with cosmological gravitation, that is, $\alpha$ is universal and does not depend on the type of atomic clock. Thus, ``null redshift experiments'' in Fig.~3 in Ref.~\cite{Will_2014} cannot be used to determine the universal cosmological correction $\alpha$. Therefore, if now we analyze the data of all other experiments in Fig.~3 in Ref.~\cite{Will_2014},  we see that these experiments allow, in principle, a value of $\alpha >10^{-3}$--$10^{-4}$. In the very recent papers Refs.~\cite{Delva_2018,Herrmann_2018}, the presented experimental results allow, in principle, of $\alpha >10^{-5}$ and they do not still contradict to our predictions Eqs.~(\ref{limit}),(\ref{cosm}).

It is also of great interest to analyze experiments in the recent paper \cite{Katori_2016}. Indeed, if we look at Fig.~3a in Ref.~\cite{Katori_2016}, we clearly see some imbalance between geodetic measurements of the geopotential difference (see red shading in Fig.~3a in Ref.~\cite{Katori_2016}) and measurements of frequency differences between clocks (see blue shaded region in Fig.~3a in Ref.~\cite{Katori_2016}). If we interpret this imbalance as the presence of $\alpha\neq 0$ in Eq.~(\ref{Dw_gen}), then we can admit that $\alpha > 10^{-3}$ ($\alpha\approx 3\times 10^{-3}$ for the midline of the blue shaded region in Fig.~3a in Ref.~\cite{Katori_2016}). It corresponds to the cosmological gravitational potential $\Phi_\text{CP}$, which is three orders of magnitude greater than the gravitational potential of the Galaxy at the Sun's orbit ($|\varphi^{}_\text{Galaxy}/c^2|\sim 10^{-6}$).

However, in our opinion, experiments on measuring of $\alpha$ in formula (\ref{Dw_gen}) should be made with two identical atomic clocks located at different heights, but at the same geographical point. This allows us: a) to minimize the time-dependent influence of the gravitational potentials of the Sun and Moon; b) to measure $[\omega({\bf r}_1)-\omega({\bf r}_2)]/\omega({\bf r}_1)$ and $[\varphi({\bf r}_1)-\varphi({\bf r}_2)]$ with maximal accuracy.

Let us consider the case where the height difference between two identical atomic clocks is $h$, that is, for points $z_1$ and $z_2=z_1+h$. For this we rewrite Eq.~(\ref{Dw_gen}) in the following form:
\begin{equation}\label{alphameter}
\frac{\Delta\omega}{\omega}-\frac{\Delta \varphi}{c^2}=\alpha\frac{\Delta \varphi}{c^2}\,.
\end{equation}
The change in the gravitational potential between the two atomic clocks is defined as:
\begin{equation}\label{DU}
 \Delta \varphi=\varphi(z_1)-\varphi(z_2)=-\int_{z_1}^{z_1+h} g(z)dz\,,
\end{equation}
where $g (z) $ is the local free fall acceleration at a point with the vertical coordinate $z$. In the case of small $h$, we can use approximation: $\Delta \varphi\approx -hg(z_1)$.

Let us estimate the possibility of measuring $\alpha$ using Eq.~(\ref{alphameter}) if the relative uncertainty of the atomic clocks is $10^{-18}$ and the height difference between the two identical atomic clocks is $h=10$~m when $\Delta\varphi/c^2\approx 1.1\times 10^{-15}$. It is obvious that the measurement error of $\Delta\omega/\omega$ corresponds to the uncertainty of the atomic clocks, that is, $10^{-18}$. Because the free fall acceleration, $g$, can be measured by modern gravimeters (for example, quantum gravimeters based on atomic interferometry or classical ballistic gravimeters) with relative accuracy much better than $10^{-6}$, the main error in determining $\Delta \varphi/c^2$ is connected with the measurement of the height $h$. If the relative position of work zones (i.e. places where atoms or ions are localized) of the interrogated atomic clocks is known with an accuracy of 1~cm, for $h=10$~m this results in an uncertainty of $10^{-18}$ for $\Delta \varphi/c^2$ in Eq.~(\ref{alphameter}). As a result, with the expression (\ref{alphameter}), $\alpha$ can be measured if $|\alpha \Delta\varphi/c^2|>10^{-18}$, that is, if $|\alpha |>10^{-3}$. The sensitivity of the measurement method
will increase ten times if the atomic clocks have a relative uncertainty of $\sim 10^{-19}$ and a measurement accuracy of $h$ of 1~mm. In this case, we can measure $|\alpha |>10^{-4}$ (for $h=10$~m).

Using similar estimates for the height $h=100$~m, we find that for atomic clocks with a relative uncertainty of $10^{-18}$ and a measurement accuracy of $h$ of 1cm (that is, $\Delta h/h\sim 10^{-4}$), one can measure $\alpha$ if $|\alpha|>10^{-4}$. For clocks with an uncertainty of $10^{-19}$ and a measurement accuracy of $h$ of 1~mm, one can measure $\alpha$ if $|\alpha |>10^{-5}$. In the same way, the reasoning can be continued for $h=1000$~m when
$|\alpha |>10^{-5}$--$10^{-6}$ can be measured. Schemes for measuring $\alpha$ with the use of spacecrafts can also be developed.

Note that the validation criterion for measurements of the cosmological correction $\alpha$ is obvious: experiments at various geographical points and for clocks with various atomic transitions should show the same results.

\section*{Conclusion}
In this paper, a hypothesis that the cosmological gravitational potential can be measured with the use of high-precision atomic clocks has been proposed and substantiated in the frame of nonmetric theory of gravitation. It was assumed that the cosmological potential is formed by all matter of the Universe (including dark matter and dark energy). The information on cosmological gravitation is contained in the nonzero cosmological correction $\alpha$ in Eq. (2). In the case of experimental confirmation of our hypothesis, a new unique window will be opened for the study of the Universe and the testing of cosmological models. In particular, if $\alpha >2\times 10^{-6}$, then it can be argued that the cosmological gravitational contribution dominates the Galaxy's gravitational potential. Further continuous monitoring for the purpose to detect possible variations of $\alpha$ can lead (if such variations exist) to the discovery of new cosmological processes, detection of dark matter (see comment \cite{comm_dm}) and gravitational waves. These results will also allow us to increase the accuracy of measurements in relativistic geodesy, chronometric gravimetry, global navigation systems, the global network of atomic clocks, etc.

If the experiments will show the opposite results, $|\alpha|\ll 10^{-6}$, this will be one of the most convincing tests for the general theory of relativity, since in this theory there is no possibility of such cosmological gravimetry (i.e., $\alpha =0$    independently of existence/absence of all other celestial bodies in the Universe). For example, in the recent papers Refs.~\cite{Delva_2018,Herrmann_2018}, the presented experimental results allow, in principle, of $\alpha >10^{-5}$ and they do not still contradict to our predictions ($\alpha >10^{-6}$). Thus, the further verifying experiments become very important.

We thank N. Ashby, C. Oates, C. Tamm, U. Sterr, A. Derevianko, M. S. Zhan, and E. M. Rasel for useful discussions and comments.

%-------------------------- BIBLIOGRAPHY ------------------------------------


\begin{thebibliography}{30}
%
\bibitem{Schioppo_2017}
M. Schioppo, R. C. Brown, W. F. McGrew, N.~Hinkley, R.~J.~Fasano, K.~Beloy, T.~H.~Yoon,
G.~Milani, D.~Nicolodi, J.~A.~Sherman, N.~B.~Phillips, C.~W.~Oates, and A.~D.~Ludlow, Nature Photonics {\bf 11}, 48–52 (2017).
%
\bibitem{Ludlow_2015}
A. D. Ludlow, M. M. Boyd, J. Ye, E. Peik, and P. O. Schmidt,
Rev. Mod. Phys. {\bf 87}, 637 (2015).
%
\bibitem{Angstmann_2004}
E. J. Angstmann, V. A. Dzuba, and V. V. Flambaum, arXiv:physics/0407141v1 (2004).
%
\bibitem{Karshenboim_2005}
S. G. Karshenboim, V. V. Flambaum, and E. Peik, {\em Handbook
of Atomic, Molecular and Optical Physics} (Springer, New York),
p. 455. (2005).
%
\bibitem{Reinhold_2006}
E. Reinhold, R. Buning, U. Hollenstein, A. Ivanchik, P. Petitjean,
and W. Ubachs, Phys. Rev. Lett. {\bf 96}, 151101 (2006).
%
\bibitem{Flambaum_2007}
V. V. Flambaum, and M. G. Kozlov, Phys. Rev. Lett. {\bf 98},
240801 (2007).
%
\bibitem{Flambaum_2009}
V. V. Flambaum, and V. A. Dzuba, Can. J. Phys. {\bf 87}, 25 (2009).
%
\bibitem{Tilburg_2015}
K. Van Tilburg, N. Leefer, L. Bougas, and D. Budker, Phys. Rev. Lett. {\bf 115}, 011802 (2015).
%
\bibitem{Derevianko_2014}
A. Derevianko and M. Pospelov,  Nature Phys. {\bf 10}, 933-936 (2014).
%
\bibitem{Stadnik_2016}
Y. V. Stadnik and V. V. Flambaum, Phys. Rev. A {\bf 93}, 063630 (2016).
%
\bibitem{Roberts_2017}
B. M. Roberts, G. Blewitt, C. Dailey, M.~Murphy, M.~Pospelov,
A.~Rollings, J.~Sherman, W.~Williams, and A.~Derevianko, arXiv:1704.06844 [hep-ph] (2017).
%
\bibitem{Einstein_1916}
A. Einstein, Annal. Physik {\bf 49}, 769 (1916)
%
\bibitem{Chou_Sc_2010}
C. W. Chou, D. B. Hume, T. Rosenband, and D.~J.~Wineland, Science {\bf 329}, 1630-1633 (2010).
%
\bibitem{Katori_2016}
T. Takano, M. Takamoto, I. Ushijima, N.~Ohmae, T.~Akatsuka,
A.~Yamaguchi, Y.~Kuroishi, H.~Munekane, B.~Miyahara,
and H.~Katori, Nature Photonics {\bf 10}, 662 (2016).
%
\bibitem{Vermeer_1983}
M. Vermeer, {\em Chronometric levelling}, Rep. Finnish Geodetic Inst. {\bf 83}, 1-7 (1983).
%
\bibitem{Nordtvedt_1975}
K. Nordtvedt, Phys. Rev. D {\bf 11}, 245 (1975).
%
\bibitem{Yudin_2017}
V. I. Yudin and A. V. Taichenachev, Laser Phys. Lett. {\bf 15}, 035703 (2018).
%
\bibitem{Wolf_2016}
P. Wolf and L. Blanchet, Class. Quantum Grav. {\bf 33}, 035012 (2016).
%
\bibitem{comm_C0}
Note that from the formal mathematical viewpoint under determination of the general gravitational potential in Eq.~(\ref{U_structure}) we can add some constant $C_0$ (universal for all Universe), which can be incorporated in cosmological potential: $(\Phi_\text{CP}({\bf r})+C_0)\rightarrow \Phi_\text{CP}({\bf r})$. In this case, cosmological potential can be positive: $\Phi_\text{CP}({\bf r}^{}_\text{E})>0$ if $C_0 >0$. However, from the physical viewpoint the existence of $C_0\neq 0$ looks quite artificially and groundlessly, because it leads to the existence of gravitational potential even for empty space in the absence of a matter.
%
\bibitem{Will_2014}
C. M. Will, Living Rev. Relativity {\bf 17}, 4 (2014).
%
\bibitem{Delva_2018}
P. Delva, N. Puchades, E. Sch$\ddot{\rm o}$nemann, F. Dilssner, C. Courde, S. Bertone, F.~Gonzalez, A. Hees, Ch.~Le~Poncin-Lafitte, F.~Meynadier, R. Prieto-Cerdeira, B. Sohet, J. Ventura-Traveset, and P. Wolf, Phys. Rev. Lett. {\bf 121}, 231101 (2018).
%
\bibitem{Herrmann_2018}
S. Herrmann, F. Finke, M. L$\ddot{\rm u}$lf, O. Kichakova, D. Puetzfeld, D. Knickmann, M. List, B. Rievers, G. Giorgi, Ch. G$\ddot{\rm u}$nther, H. Dittus, R. Prieto-Cerdeira, F. Dilssner, F. Gonzalez, E. Sch$\ddot{\rm o}$nemann, J. Ventura-Traveset, and C. L$\ddot{\rm a}$mmerzahl, Phys. Rev. Lett. {\bf 121}, 231102 (2018).
%
\bibitem{comm_dm}
Indeed, it is possible to assume that due to the motion of the Sun in the Galaxy the trajectory of this motion will cross areas with spatially non-uniform distribution of dark matter, and, respectively, with spatially non-uniform gravitational potential. In this case, there will be some variations of the cosmological correction $\alpha$, which can be experimentally detected.




\end{thebibliography}
\end{document}